\documentclass[rmp,twocolumn, preprintnumbers,amsmath, showpacs, amssymb, superscriptaddress, byrevtex]{revtex4}
\usepackage{graphicx}
\usepackage{dcolumn}

\begin{document}
\title{Photoluminescence from localized states in disordered indium nitride}
\author{Bhavtosh Bansal}
\email{bhavtosh.bansal@science.ru.nl}  \affiliation{IMM, High Field Magnet Laboratory
HFML, University of Nijmegen, Toernooiveld 7, 6525 ED Nijmegen,
The Netherlands}\affiliation{INPAC-Institute
for Nanoscale Physics and Chemistry, Pulsed Fields Group,
Katholieke Universiteit Leuven, Celestijnenlaan 200D, Leuven
B-3001, Belgium}
\author{Abdul Kadir}
 \affiliation{Tata Institute of Fundamental Research, 1 Homi Bhabha Road, Mumbai, India}
\author{Arnab Bhattacharya}
 \affiliation{Tata Institute of Fundamental Research, 1 Homi Bhabha Road, Mumbai, India}
\author{V. V. Moshchalkov}
\affiliation{INPAC-Institute for Nanoscale Physics and Chemistry,
Pulsed Fields Group, Katholieke Universiteit Leuven,
Celestijnenlaan 200D, Leuven B-3001, Belgium}
\date{\today}
\begin{abstract}
Photoluminescence spectra from disordered InN were studied in very high magnetic fields. The samples had Gaussian spectra with low temperature emission peaks at 0.82 and 0.98eV respectively. The average spatial extent of the excitonic wave functions, inferred from the diamagnetic shift, is only 2-3nm. This shows that the recombination is from an ensemble of highly localized states within a landscape of a smooth (classical) disorder potential of strength of the order of 10meV. The anomalies in the temperature dependence of the photoluminescence peak and linewidth give further support to the picture of trapped photoexcited carriers.
\end{abstract}
\maketitle
After a few years of active controversy, there is finally consensus that the value of energy gap of bulk indium nitride is about $0.7$ eV[\onlinecite{davydov}, \onlinecite{wu}]. While it is now generally agreed that the experimentally observed variation in the energy gap across samples can be accounted for by the carrier induced band filling (Burstein-Moss shift)[\onlinecite{wu}], and that InN has all the properties of a direct gap semiconductor, so far there are only a few experimental investigations that directly probe its band structure [\onlinecite{butcher_localized states nature}] or the nature of the electronic states[\onlinecite{Inushima, sandip}]. Interpretation of most of the usual optical and transport measurements[\onlinecite{jena}] is based on gross averages, where the information about the nature of the electronic wave functions is buried deep within the integrals over the electronic distribution function and the density of states.

 InN has been notoriously difficult grow and the observed physical properties are often dominated by the defects and impurities present in the material. Specifically, in the moderately high carrier concentration range ($5\times 10^{18}$--$5\times 10^{19}$ cm$^{-3}$), the origin of optical emission is still unclear. Magneto-transport experiments[\onlinecite{Inushima}] have revealed an anomalous band structure for InN above the carrier concentration of $5\times 10^{18}$ cm$^{-3}$. The signature of anomalies was also observed in the temperature-dependent photoluminescence (PL) measurements on samples with carrier densities of this order[\onlinecite{Wu bandgap tempdp}, \onlinecite{localization InN}]. The material also exhibits superconductivity that is most probably due to indium oxide contamination[\onlinecite{kadir superconductivity}].

A characteristic that distinguishes a band-edge state from a defect state is---by definition---the spatial extent[\onlinecite{stefani_mature mano}] of wave functions. Band-edge emission is either excitonic or due to free-carrier recombination; the electronic wave functions in either case are extended. The defect levels on the other hand are characterized by a small wave function. By analyzing the PL spectra from InN samples in very high magnetic fields, this letter attempts to determine the spatial extent of the wave function of the states participating in light emission.

\noindent
{\em Experimental.---}The wurtzite InN was grown by metal-organic vapor phase epitaxy on sapphire substrates with a GaN buffer at the temperature of 530 $^o$C. The details of growth, structural, optical and electrical characterization have been reported in references \onlinecite{kadir_jcg} and \onlinecite{kadir_apl}. Here we describe the results from two samples, labeled {\em A} and {\em B}. {\em Sample A} had a carrier density of $1.7\times 10^{19}$cm$^{-3}$ and a mobility of 195 cm$^{2}$/V-s and {\em Sample B} had a carrier density of $3.0\times 10^{19}$cm$^{-3}$ and a mobility of 59 cm$^{2}$/V-s. X-ray diffraction measurements showed that the samples were single-crystalline, with a residual hydrostatic strain[\onlinecite{kadir_apl}]. PL measurements were performed in pulsed magnetic fields of up to 50 tesla at liquid helium temperatures. Samples were non-resonantly excited with a frequency-doubled diode-pumped Nd:YAG laser emitting at 532 nm. Multiple spectra were recorded during a single pulse on a fast InGaAs linear array detector. The detector has a sharp cut-off beyond 1600nm and hence the long-wavelength measurement range and the choice of the samples was limited to this value. The system response was calibrated against a standard black-body source and the spectra were corrected for it to rule out any error from the detector cut-off[\onlinecite{cutoff}].

Fig.1 shows the representative PL spectra (along with the fit to the Gaussian waveforms) from the two samples used for magneto-PL analysis. The integrated intensity from the sample with a higher doping ({\em Sample A}) is about a factor of two smaller indicating a greater role of defects.

Fig.2 shows the location of the PL peak energy (inferred from the Gaussian fits) of the two samples as a function of magnetic field. Due to relatively poor luminescence and the large linewidths, there was considerable scatter in the data, especially from {\em Sample A}. (But also note that the range of the y-axis in Fig.2 is only about 4meV.) The most striking observation is the extremely small magnitude of the diamagnetic shift, just about 2meV for {\em Sample A} and 2.5 meV for {\em Sample B}, even in fields up to about 50 tesla. The lineshape, the linewidth, and the emission intensity did not noticeably change up to the highest magnetic field.

\begin{figure}
\includegraphics[width=90mm]{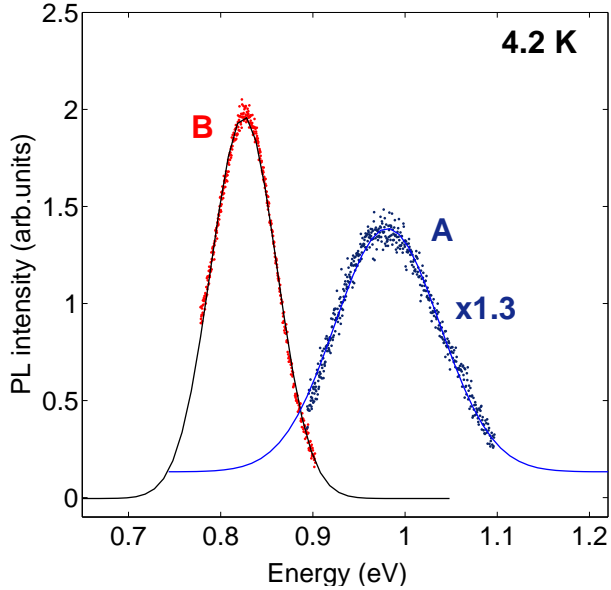}
\caption{(color online) Typical PL spectra from the two samples, {\em Sample A} and {\em Sample B}, measured at 4.2K which were used in the analysis. The InGaAs array detector has a sharp cut-off below around 0.78 eV. Since the fits to the Gaussian waveform (solid lines) is very good, the measurement accuracy is not limited by our inability to capture the complete emission spectra.}
\end{figure}
The solid lines in Fig. 2 are fits to equation $E(B)=E_0+\Gamma B^2$, where $E_0$ is the zero field emission energy and $\Gamma$ is a constant. Since the shift for both the samples is quadratic in the entire field range, the states contributing to emission are bound states of zero angular momentum (S-orbital-like). Secondly the absence of a linear region even beyond 40 tesla implies that the samples are still in the `low-field regime' (exciton size $<$ magnetic length) and the second-order perturbation theory (that predicts a quadratic diamagnetic shift of the spectrum with the magnetic field) is a good description[\onlinecite{footnote1}]. Thus we have an upper bound on the exciton size without making any assumptions about the exciton mass. The magnetic length at 40 tesla is about 4 nm.
\begin{figure}
\includegraphics[width=90mm]{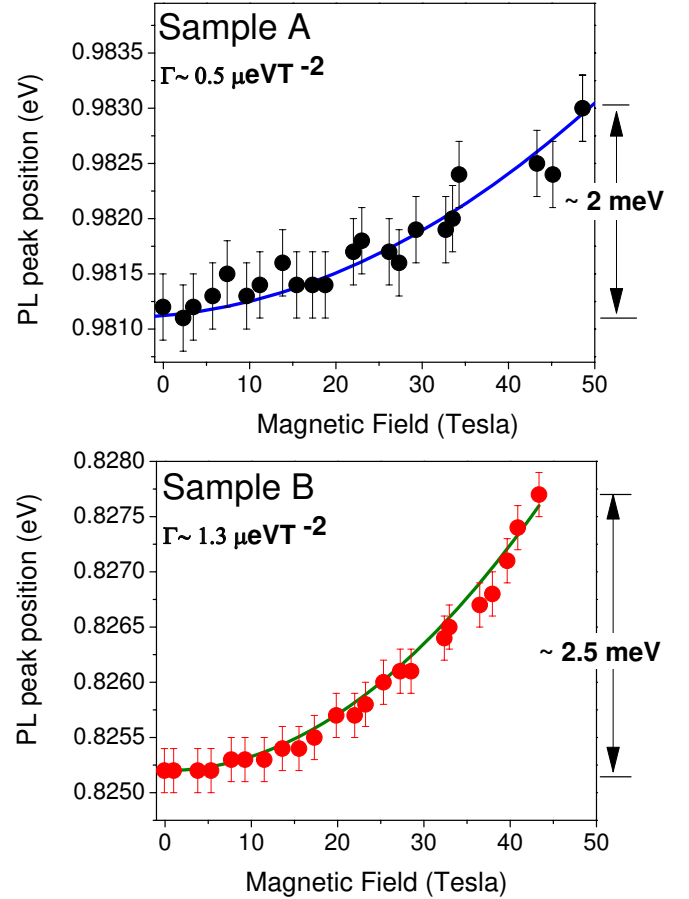}
\caption{(color online)(circles) Diamagnetic shift at 4.2K of {\it Sample A} and {\it Sample B} averaged over many data points. Error bars are indicative of the scatter in the raw data. (solid line) Fit to a quadratic function, $E(B)=E_0+\Gamma B^2$. Note the absolute magnitude of the shift for both the samples is very small.}
\end{figure}

The diamagnetic shift $ E(B)$ is quantitatively governed by the following relationship:
\begin{eqnarray}{\label{eqn1}}
E(B)-E_0={q^2\langle\rho^2\rangle\over 8 \mu}B^2 \equiv \Gamma B^2
\end{eqnarray}
Thus the diamagnetic shift coefficient $\Gamma$
explicitly depends on the average exciton $\sqrt{\langle\rho^2\rangle}$ and the reduced effective mass $\mu$. $q$ is the electronic charge and $B$ denotes the magnetic field.

Fitting the data in Fig. 2 to the above equation gives $\Gamma=0.5\mu eV T^{-2}$ and $1.3 \mu eV T^{-2}$ for {\em Sample A} and {\em Sample B} respectively. Taking the value[\onlinecite{wu}] of the reduced effective mass $\mu$ to be 0.1$m_0$, close to the literature value for the electron mass in InN at a carrier density of $1\times 10^{19}$, we find that the effective size of the light emitting state $\sqrt{\langle\rho^2\rangle}$ is only $1.5$ nm for {\em Sample  A} and $2.4$ nm for {\em Sample B}.

Contrast this to the value of the excitonic Bohr radius $a_B$ expected from the hydrogenic exciton model where $a_B=4\pi\epsilon\epsilon_0\hbar^2/(\mu e^2)$. With $\mu=0.1$ and the static dielectric constant $\epsilon=15.3$, the expected excitonic radius in InN should be $8$ nm. Therefore such a small value of the size of the light emitting state goes against the hypothesis that the emission around $0.8$eV is from the {\em usual} Wannier exciton-like excitations. On the other hand, the emission state radius of about 2nm also rules out deep-defect emission since such states are expected to be extended over one or two atomic sites. In this case, the PL typically shows no diamagnetic shift[\onlinecite{stefani_mature mano}]. This leads us to the conclusion that the electronic states participating in light emission are disorder-localized states whose spatial extent  is intermediate between the free exciton size and the interatomic spacing.

On dimensional grounds, the energy required to localize an exciton[\onlinecite{runge}] is approximately $E_\textrm{loc}=\hbar^2/[2 \mu l_c^2]$, where the localization length $l_c$ can be identified with the Bohr radius inferred from the magneto-PL measurements. For $l_c=2$nm and the effective mass $\mu=0.1m_0$, $E_\textrm{loc}$ turns out to be about $10$ meV. Interestingly, the same order of confinement energies were found in reference \onlinecite{localization InN} through three parameter fit to the temperature dependence of the PL intensity.
Comparing these to the values of the linewidths $\sigma\sim 100$meV, we find that $\sigma/E_\textrm{loc}\gg1$. This implies a classical limit where the potential is locally smooth and optically active states look like $k=0$ plane waves[\onlinecite{runge}]. The fact that the spectra in Fig.1 could be nicely to fitted to Gaussian waveforms confirms that the localization is of classical origin. Interference effects related to Anderson localization are relatively unimportant as they would have resulted in a skewed lineshape.
\begin{figure}[t]
\includegraphics[width=70mm]{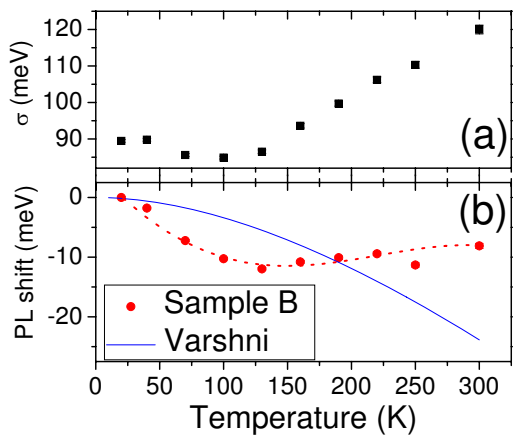}
\caption{(color online) Temperature dependence of the (a) linewidth and (b) shift in the peak PL energy for {\it Sample B}. Non-monotonic behavior of both indicates a strong role of carrier localization. (b) Solid line (labeled Varshni) is the shift expected from the intrinsic bandgap change and dotted line is just to guide the eye. }
\end{figure}

Such classical localization predicts that the emission and the absorption spectra should have a Stokes' shift that is related to the linewidth[\onlinecite{linewidthfootnote}]. For a Gaussian density of states, the Stokes' shift is of the order of the linewidth $\sigma$ at `low' temperatures and is given by $\sigma^2/[k_BT]$ in the high temperature limit[\onlinecite{runge, bhavtosh stokes,trapping vs thermalization}]. The crossover temperature separating these limits is itself of the order of $\sigma$, which in the present case would be more than 1000K. The room temperature absorption spectra measured on sample B shows an energy gap of 0.94 eV. Indeed the Stokes' shift ($\sim$ 130meV) is of the order of the linewidth at 300K (see Fig.4).

Fig.4 shows the temperature dependence of the linewidth and the shift in the PL peak position for {\it Sample B}. Both are non-monotonic. As was previously found by Wu, et al.[\onlinecite{Wu bandgap tempdp}], the temperature dependence of the PL peak is sample-dependent and does not follow the change in the bandgap (depicted here by the solid line with the Varshni parameters taken from reference [\onlinecite{inn review}]). The shift is much smaller, corresponds to local redistribution (partial thermalization[\onlinecite{bhavtosh thermalization}]) of carriers with increasing temperature, and is largely uncorrelated to the change in the intrinsic bandgap value. Such behavior is often seen in quantum dots [\onlinecite{bhavtosh thermalization}] and nitride-based semiconductor alloys such as GaInN[\onlinecite{eliseev}] and dilute nitrides [\onlinecite{bhavtosh stokes}].
Clearly the same physics is at work here. In the present case, we have a smoothly varying potential energy landscape where the variations are caused by root mean squared fluctuations in the screened coulombic potential of the ionized dopant atoms[\onlinecite{Arnaudov, localization InN}], as well as other compositional (indium oxide[\onlinecite{kadir superconductivity}]) and topological disorder (nitrogen vacancies[\onlinecite{kadir_apl}]) in the sample. In context of GaInN, localized states may in fact enhance the radiative efficiency and could be the reason why these nitride-based light emitting diodes are technologically viable[\onlinecite{odennell}], despite the large density of line dislocations and other defects.

\noindent
{\em Summary.---}
We have attempted to clarify the nature of light emitting states in disordered InN. Specifically, the high field magneto-PL was used to unambiguously prove that the PL emission at 0.98 eV and 0.82 eV, from the two InN samples studied here, is from localized states whose size is of the order of 2nm. The localization was found to be of classical origin. The values of the Stokes' shift and the linewidths, and the behavior of the temperature dependence of PL suggested that these trapped states are analogous to those found in other highly disordered semiconductors.

\noindent
We thank Dr. Sandip Ghosh and Dr. Antonio Polimeni for insightful comments on the manuscript.
B.B. acknowledges the support of the DeNUF Project of the European Commission FP6 Contract 01176.

\end{document}